\documentclass[A4]{article}
\usepackage[margin=2cm]{geometry}
\usepackage{amsmath}
\usepackage{amssymb}
\usepackage{amsfonts}
\usepackage{graphicx,color}
\usepackage{float}
\usepackage{rotating}
\usepackage{tikz}

\newcounter{NN}
\newtheorem{theorem}[NN]{Theorem}

\newtheorem{conjecture}[NN]{Conjecture}

\def\ut#1{\underset{\text{{\tiny $\sim$}}}{#1}}
\def\uh#1{\underset{\text{{\tiny \rotatebox{90}{$\rangle$}}}}{#1}}
\def\ud#1{\underset{\cdot}{#1}}

\def\P{P}
\def\Q{Q}
\def\R{R}
\def\v{{X}}
\def\w{{Y}}
\def\V{{V}}
\def\W{{W}}

\begin{document}
\title{Duality for discrete integrable systems II}
\author{Peter H.~van der Kamp$^{1,3}$, G.R.W Quispel$^1$, Da-jun Zhang$^2$}
\date{$^1$Department of Mathematics and Statistics\\
La Trobe University, Victoria 3086, Australia\\
$^2$Department of Mathematics,
Shanghai University \\
Shanghai 200444,
China\\
[5mm]
$^3$ Corresponding author, P.vanderKamp@LaTrobe.edu.au
}
\maketitle

\begin{abstract}
We generalise the concept of duality to lattice equations. We derive a novel 3 dimensional lattice equation, which is dual to the lattice AKP equation. Reductions of this equation include Rutishauser's quotient-difference (QD) algorithm, the higher analogue of the discrete time Toda (HADT) equation and its corresponding quotient-quotient-difference (QQD) system, the discrete hungry Lotka-Volterra system, discrete hungry QD, as well as the hungry forms of HADT and QQD. We provide three conservation laws, we conjecture the equation admits N-soliton solutions and that reductions have the Laurent property and vanishing algebraic entropy.
\end{abstract}

\begin{center}{\small
}
\end{center}

\section{Introduction}
Our aim in the present paper is threefold:
\begin{enumerate}
\item To generalise the concept of duality (introduced in \cite{QCR} for ordinary difference equations) to lattice equations;
\item To use duality to derive the 3 dimensional (3D) lattice equation
\begin{equation} \label{DAKP}
\begin{split}
0&= a_{{1}} \left(\tau_{{k-1,l+1,m+1}}\tau_{{k+1,l,m}}\tau_{{k+1,l,m+1}}\tau_{{k+1,l+1,m}}
-\tau_{{k,l,m+1}}\tau_{{k,l+1,m}}\tau_{{k,l+1,m+1}}\tau_{{k+2,l,m}} \right)\\
&\ \ \ + a_{{2}} \left(\tau_{{k,l+1,m}}\tau_{{k,l+1,m+1}}\tau_{{k+1,l-1,m+1}}\tau_{{k+1,l+1,m}}
-\tau_{{k,l,m+1}}\tau_{{k,l+2,m}}\tau_{{k+1,l,m}}\tau_{{k+1,l,m+1}} \right)\\
&\ \ \ + a_{{3}} \left( \tau_{{k,l,m+1}}\tau_{{k,l+1,m+1}}\tau_{{k+1,l,m+1}}\tau_{{k+1,l+1,m-1}}
-\tau_{{k,l,m+2}}\tau_{{k,l+1,m}}\tau_{{k+1,l,m}}\tau_{{k+1,l+1,m}} \right)\\
&\ \ \ + a_{{4}} \left( \tau_{{k,l,m}}\tau_{{k,l+1,m+1}}\tau_{{k+1,l,m+1}}\tau_{{k+1,l+1,
m}}-\tau_{{k,l,m+1}}\tau_{{k,l+1,m}}\tau_{{k+1,l,m}}\tau_{{k+1,l+1,m+1}} \right).
\end{split}
\end{equation}
\item To provide conservation laws for equation (\ref{DAKP}), to present reductions to two dimensional integrable systems, and to support our conjecture that equation (\ref{DAKP}) admits $N$-soliton solutions and its reductions have the Laurent property and vanishing algebraic entropy.
\end{enumerate}
Most of currently known integrable 3D lattice equations are related to discretizations of the three continuous 3D Kadomtsev-Petviashvili equations called AKP, BKP and CKP. The lattice AKP equation,
\begin{equation} \label{AKP}
A\tau_{k+1,l,m}\tau_{k,l+1,m+1}+
B\tau_{k,l+1,m}\tau_{k+1,l,m+1}+
C\tau_{k,l,m+1}\tau_{k+1,l+1,m}=0,
\end{equation}
was first derived by Hirota \cite{Hir}, and is also called the Hirota-Miwa equation \cite{Miw}. The more general lattice BKP equation (also called the Miwa equation),
\begin{equation} \label{BKP}
A\tau_{k+1,l,m}\tau_{k,l+1,m+1}+
B\tau_{k,l+1,m}\tau_{k+1,l,m+1}+
C\tau_{k,l,m+1}\tau_{k+1,l+1,m}+
D\tau_{k,l,m}\tau_{k+1,l+1,m+1}=0,
\end{equation}
was first found by Miwa in \cite{Miw}).
The lattice CKP equation,
\begin{align}
&(\tau_{k,l,m}\tau_{k+1,l+1,m+1}+\tau_{k+1,l,m}\tau_{k,l+1,m+1}-\tau_{k,l+1,m}\tau_{k+1,l,m+1}-\tau_{k,l,m+1}\tau_{k+1,l+1,m})^2\nonumber\\
=&4(\tau_{k,l,m}\tau_{k+1,l,m+1}-\tau_{k,l+1,m}\tau_{k,l,m+1}) (\tau_{k+1,l,m}\tau_{k+1,l+1,m+1}-\tau_{k+1,l+1,m}\tau_{k+1,l,m+1})
\label{CKP}
\end{align}
was first derived by Kashaev as a 3D lattice model associated with the local Yang-Baxter relation \cite{Kas},
and later was independently found by King and Schief \cite{KS}
as a superposition principle for the continuous CKP equation.
This equation is also formulated as a hyperdeterminant in \cite{TW}.

\begin{figure}[h!]
\begin{center}
\begin{tikzpicture}[rotate=0]
\fill \foreach \p in {(0, 0), (0., 1.20), (0., 2.40), (-.6, -.6), (-1.2, -1.2), (1.3, -.3), (2.6, -.6), (.7, .30), (-.6, .60), (-1.9, .90), (1.3, .90), (1.9, 1.50), (.7, -.90), (.7, -2.10)} {\p circle (3pt)};
\draw[thick]     (-1.2, -1.2)--(0, 0)--(0., 2.40)
             (-1.9, 0.90)--(0.7, 0.30)--(1.9, 1.50)
  (0, 0)--(1.3, -0.3)--(1.3, 0.90)--(0., 1.20)--(-0.6, 0.60)--(-0.6, -0.6)--(0.7, -0.90)--(1.3, -0.3)--(2.6, -0.6)
                   (0.7, -2.10)--(0.7, 0.30)
;
\end{tikzpicture}
\caption{\label{STC} The 14-point stencil of equation (\ref{DAKP}).}
\end{center}
\end{figure}
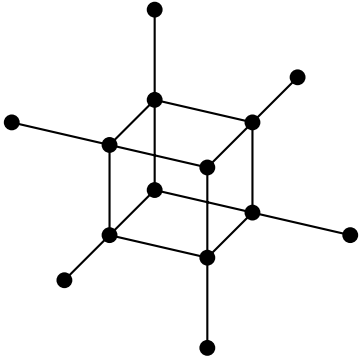

The AKP equation is a bilinear equation on a six-point octahedral stencil ($A_3$ lattice). Equations of this type have been classified with respect to multi-dimensional consistency in \cite{ABS}. The lattice BKP and CKP equations are both defined on an 8-point cubic stencil. However, whereas lattice BKP is bilinear, the lattice CKP is quartic and nonlinear. A nonlinear form of the AKP equation (quartic and defined on a 10-point stencil) was given in \cite[equation 5.5]{GRPSW}. A quintic nonlinear non-potential lattice AKP equation was given in \cite[equation 3.19]{FuNi}. This equation is defined on a 10-point stencil \cite[Figure 3]{FuNi}. A quadrilinear 3D lattice equation related to the lattice BKP equation, defined on a 14-point stencil ($D_3$ lattice), is presented in \cite[Equation 24]{KS}. Our equation (\ref{DAKP}), which we will obtain as a dual to the AKP equation (\ref{AKP}), is a quadrilinear equation defined on the 14 point stencil depicted in Figure \ref{STC}.
To our knowledge equation (\ref{DAKP}) is new.  Given that the number of known integrable 3D lattice equations is quite small, cf. \cite[Sections 3.9-3.10]{HJN}, any possible addition to this number would seem worthwhile pursuing.

The idea of duality for ordinary difference equations is as follows: given an ordinary difference equation (O$\Delta$E), $E=E(u_n,u_{n+1},\ldots,u_{n+d})=0$, with an integral, $K[n]=K(u_n,u_{n+1},\ldots,u_{n+d-1})$, the difference of the integral with its upshifted version factorises
$K[n+1] - K[n] = E \Lambda$. The quantity $\Lambda$ is called an integrating factor. The equation $\Lambda=0$ is a dual equation to the equation $E=0$, both equations share the same integral. If $E=0$ has several integrals $K_i$, then a linear combination of them gives rise to a dual with parameters:
\[
\sum_i a_i K_i[n+1] - \sum_i a_i K_i[n] = E \left( \sum_i a_i \Lambda_i \right).
\]
In \cite{QCR} duals to $(d-1,-1)$-periodic reductions of the modified Korteweg-de Vries (mKdV) lattice equation are shown to be integrable maps, namely level-set-dependent mKdV maps. In \cite{DTKQ} a novel hierarchy of maps is found by applying the concept of duality to the linear equation $u_{n}=u_{n+d}$, and $\lfloor \frac{d-1}{2}\rfloor$ integrals are provided explicitly. The integrability of these maps is established in \cite{HKQ}. We note that dual equations are not necessarily integrable, examples exist where the dual is not integrable \cite{DKQ}. In \cite{JV}, the authors study several integrable 4th order maps and integrable maps that are dual to them.

Given a 2D lattice equation, $E=E(u_{k,l},\ldots,u_{k+d,l+e})=0$, instead of considering differences of integrals we now consider conservation laws:
\[
P[k+1,l] - P[k,l] + Q[k,l+1] - Q[k,l] = E \Lambda.
\]
Here the quantity $\Lambda$ is called the characteristic of the conservation law. Again the equation $\Lambda=0$ or a linear combination,
$\sum_i a_i \Lambda_i =0$, can be viewed as the dual equation to $E=0$. The situation for 3D lattice equations is similar.

The structure of the paper is as follows. In section \ref{sdAKP} we present a 3D lattice equation which is dual to the lattice AKP equation, and we provide a matrix formula which simultaneously
captures four conservation laws for the AKP equation as well as three conservation laws for the dual AKP equation. In section \ref{scqds} we show that these conservation
laws give rise to quotients-difference formulations, in the same way that Rutishauser's quotient-difference (QD) algorithm \cite{Rut} is a quotient-difference formulation of the discrete-time Toda equation \cite{Hir2}.
In section \ref{stnss} we provide 1-soliton and 2-solition solutions, and we provide evidence to support a conjectured form for the general $N$-soliton solution. In section \ref{sLp} we provide evidence to support our conjecture that 2-periodic reductions to ordinary difference equations have the Laurent property. In section \ref{DG} we provide details of calculations which indicate that 2-periodic reductions of the dual AKP equation have quadratic growth.
In section \ref{srt2dle} we show that reductions of the dual AKP equation (\ref{DAKP}) to 2D lattice equations include the higher analogue of the discrete time Toda (HADT) equation and its corresponding quotient-quotient-difference (QQD) system \cite{SNK}, the discrete hungry Lotka-Volterra system, discrete hungry QD, as well as the hungry forms of HADT and QQD introduced in \cite{CCHT}.

\section{Derivation of a dual to the lattice AKP equation, and a matrix conservation law} \label{sdAKP}
Seven characteristics of conservation laws for the lattice AKP equation can be obtained from the results in \cite{MQ}.
We choose to only consider the parameter independent ones and we set all arbitrary functions equal to 1.
We will denote shifts in $k$ using tildes, shifts in $l$ by hats, and shifts in $m$ by dots, e.g.
$\underset{\dot{ }}{\hat{\tilde{\tau}}}=\tau_{k+1,l+1,m-1}$.
The four characteristics (denoted $\Lambda_1,\Lambda_2,\Lambda_3,\Lambda_7$ in \cite{MQ}) are given by
\[
\W=\left(\dfrac{\ut{\hat{\dot{\tau}}}}{\hat{\dot{\tau}}\hat{\tau}\dot{\tau}}-
\dfrac{\tilde{\tilde{\tau}}}{\tilde{\tau}\tilde{\hat{\tau}}\tilde{\dot{\tau}}},\
\dfrac{\uh{\tilde{\dot{\tau}}}}{\tilde{\dot{\tau}}\tilde{\tau}\dot{\tau}}-
\dfrac{\hat{\hat{\tau}}}{\hat{\tau}\hat{\dot{\tau}}\tilde{\hat{\tau}}},\
\dfrac{\ud{\tilde{\hat{\tau}}}}{\tilde{\hat{\tau}}\tilde{\tau}\hat{\tau}}-
\dfrac{\dot{\dot{\tau}}}{\dot{\tau}\tilde{\dot{\tau}}\hat{\dot{\tau}}},\
\dfrac{\tau}{\tilde{\tau}\hat{\tau}\dot{\tau}}-
\dfrac{\tilde{\hat{\dot{\tau}}}}{\tilde{\hat{\tau}}\hat{\dot{\tau}}\tilde{\dot{\tau}}}
\right).
\]
One can now check the following matrix conservation law
\begin{equation} \label{CONS}
\tilde{\P}-\P+\hat{\Q}-\Q+\dot{\R}-\R=\V^T\W,
\end{equation}
where $\P,\Q,\R$ are the $3\times4$ matrices
\[
\P=
\begin{pmatrix}
-\frac{\tilde{\tau}\ut{\hat{\dot{\tau}}}}{\hat{\tau}\dot{\tau}} & 0 & 0 &
-\frac{\tau\hat{\dot{\tau}}}{\hat{\tau}\dot{\tau}} \\[2mm]
-\frac{\tilde{\tau}\ut{\hat{\tau}}}{\hat{\tau}\tau} & 0 & \frac{\dot{\tau}\ud{\hat{\tau}}}{\hat{\tau}\tau} & 0 \\[2mm]
-\frac{\tilde{\tau}\ut{\dot{\tau}}}{\dot{\tau}\tau} & \frac{\hat{\tau}\uh{\dot{\tau}}}{\dot{\tau}\tau} & 0 & 0
\end{pmatrix},\quad
\Q=\begin{pmatrix}
0&-\frac{\hat{\tau}\uh{\tilde{\tau}}}{\tilde{\tau}\tau} &
\frac{\dot{\tau}\ud{\tilde{\tau}}}{\tilde{\tau}\tau} & 0 \\[2mm]
0 & -\frac{\hat{\tau}\uh{\tilde{\dot{\tau}}}}{\tilde{\tau}\dot{\tau}} & 0 & -\frac{\tau \dot{\tilde{\tau}}}{\tilde{\tau}\dot{\tau}} \\[2mm]
\frac{\tilde{\tau}\ut{\dot{\tau}}}{\dot{\tau}\tau} & -\frac{\hat{\tau}\uh{\dot{\tau}}}{\dot{\tau}\tau} & 0 & 0
\end{pmatrix},\quad
\R=\begin{pmatrix}
0 & \frac{\hat{\tau}\uh{\tilde{\tau}}}{\tilde{\tau}\tau} &
-\frac{\dot{\tau}\ud{\tilde{\tau}}}{\tilde{\tau}\tau} & 0 \\[2mm]
\frac{\tilde{\tau}\ut{\hat{\tau}}}{\hat{\tau}\tau} & 0 & -\frac{\dot{\tau}\ud{\hat{\tau}}}{\hat{\tau}\tau} & 0 \\[2mm]
0 & 0 & -\frac{\dot{\tau}\ud{\tilde{\hat{\tau}}}}{\tilde{\tau}\hat{\tau}} & -\frac{\tilde{\hat{\tau}}\tau}{\tilde{\tau}\hat{\tau}}
\end{pmatrix},
\]
and $\V^T$ denotes the transpose of
\[
\V=\left(
\hat{\dot{\tau}}\tilde{\tau},\
\tilde{\dot{\tau}}\hat{\tau},\
\hat{\tilde{\tau}}\dot{\tau}\right).
\]
Denoting two vectors of coefficients by $\v=\left( A , B , C \right)$ and $\w=\left(a_1, a_2, a_3, a_4 \right)$, we have that
$\v\V^T=0$ represents the AKP equation (\ref{AKP}) and the equation $\W\w^T=0$ is equivalent to equation (\ref{DAKP}).

Hence, pre-multiplying (\ref{CONS}) with $\v$  gives four conservation laws for the lattice AKP equation, and post-multiplying (\ref{CONS}) with $\w^T$ yields three conservation laws for equation (\ref{DAKP}). Thus the lattice AKP equation and equation (\ref{DAKP}) are dual to each other.

\section{Corresponding quotients-difference systems} \label{scqds}
The lattice AKP equation and the dual AKP equation can each be written as a system of one difference equation combined with a number of quotient equations.

Let us introduce variables
\[
p=\frac{\tau\hat{\dot{\tau}}}{\hat{\tau}\dot{\tau}},\qquad
q=\frac{\tau \dot{\tilde{\tau}}}{\tilde{\tau}\dot{\tau}},\qquad
v=\frac{\tilde{\hat{\tau}}\tau}{\tilde{\tau}\hat{\tau}}.
\]
Note that $p=-P_{14}$, $q=-Q_{24}$ and $v=-R_{34}$. Hence, the fourth conservation law for the AKP equation can be written as the difference equation
\begin{equation} \label{DEAKP}
A(\tilde{p}-p)+B(\hat{q}-q)+C(\dot{v}-v)=0.
\end{equation}
Taking logarithms, we find
\begin{align*}
\ln(p)&=c+\hat{\dot{c}}-\hat{c}-\dot{c}=(1-\hat{S})(1-\dot{S})c \\
\ln(q)&=c+\tilde{\dot{c}}-\tilde{c}-\dot{c}=(1-\tilde{S})(1-\dot{S})c \\
\ln(v)&=c+\tilde{\hat{c}}-\tilde{c}-\hat{c}=(1-\tilde{S})(1-\hat{S})c
\end{align*}
where $c=\ln(\tau)$, capital $\tilde{S}$ denotes the shift operator in $k$ (and similarly $\hat{S}$ and $\dot{S}$ represent shifts in $l$ resp. $m$), and $1$ is the identity. This gives $(1-\tilde{S})\ln(p)=(1-\hat{S})\ln(q)=(1-\dot{S})\ln(v)$, which can be written in quotient form,
\begin{equation}\label{QQ}
\frac{\tilde{p}}{p}=\frac{\hat{q}}{q}=\frac{\dot{v}}{v}.
\end{equation}
As (\ref{QQ}) contains only two independent equations the system of equations for
$p,q,v$ defined by (\ref{DEAKP}) and (\ref{QQ}) can be referred to as a QQD-system.

Similarly, we can write the dual AKP equation in variables
\begin{equation} \label{uzwv}
u=\frac{\tilde{\tau}\ut{\dot{\tau}}}{\tau\dot{\tau}},\quad
z=\frac{\hat{\tau}\uh{\dot{\tau}}}{\tau\dot{\tau}},\quad
w=\frac{\dot{\tau}\ud{\hat{\tilde{\tau}}}}{\tilde{\tau}\hat{\tau}},
\end{equation}
and the variable $v$ introduced above. We have $u=-P_{31}=Q_{31}$, $z=P_{32}=-Q_{32}$, and $w=R_{33}$. The third conservation law becomes
\begin{equation} \label{D}
a_1(\hat{u}-\tilde{u})+a_2(\tilde{z}-\hat{z})+a_3(w-\dot{w})+a_4(v-\dot{v})=0.
\end{equation}
Taking logarithms we find
\[
\ln(u)=\frac{(\tilde{S}-1)(\tilde{S}-\dot{S})}{\tilde{S}}c,\quad
\ln(z)=\frac{(\hat{S}-1)(\hat{S}-\dot{S})}{\hat{S}}c,\quad
\ln(w)=\frac{(\tilde{S}-\dot{S})(\hat{S}-\tilde{S})}{\dot{S}}c.
\]
One can now derive quotient equations which are either ratios of quadratic terms
\[
\frac{\hat{\hat{\tilde{u}}}\hat{\tilde{u}}}{\dot{\hat{\tilde{u}}}\hat{\tilde{u}}}=
\frac{\hat{\tilde{\tilde{z}}}\dot{\hat{z}}}{\dot{\hat{\tilde{z}}}\hat{\tilde{z}}}=
\frac{\dot{\hat{\tilde{w}}}\dot{w}}{\dot{\tilde{w}}\dot{\hat{w}}}=
\frac{\hat{\tilde{v}}\dot{\dot{v}}}{\dot{\tilde{v}}\dot{\hat{v}}},
\]
or ratios of linear terms
\begin{equation} \label{Q3}
\frac{\tilde{\hat{u}}}{\tilde{u}}=\frac{\tilde{v}}{\dot{v}},\quad
\frac{\tilde{\hat{u}}}{\dot{\tilde{u}}}=\frac{\dot{\tilde{w}}}{\dot{w}},\quad
\frac{\tilde{\hat{z}}}{\hat{z}}=\frac{\hat{v}}{\dot{v}},\quad
\frac{\tilde{\hat{z}}}{\dot{\hat{z}}}=\frac{\dot{\hat{w}}}{\dot{w}},
\end{equation}
of which only three are independent. In the sequel, we will refer
to the system of quotient and difference equations (\ref{D}) and (\ref{Q3}) as the Q$^3$D-system.

\section{The $N$-soliton solution} \label{stnss}
\subsection*{1-soliton}
Equation (\ref{DAKP}) admits the 1-soliton solution $\tau_{k,l,m}=1+c_1 x_1^ky_1^lz_1^m$ with dispersion relation $Q_1=0$, where
\begin{align}
Q_i=
&y_iz_i \left( x_i-1 \right)  \left( x_i-y_i
 \right)  \left( x_i-z_i \right) a_{{1}}+x_iz_i
 \left( y_i-1 \right)  \left( y_i-x_i \right)  \left( y_
i-z_i \right) a_{{2}}\notag\\
&+x_iy_i \left( z_i-1 \right)
 \left( z_i-x_i \right)  \left( z_i-y_i \right) a_{{3}
}+x_iy_iz_i \left( x_i-1 \right)  \left( y_i-1
 \right)  \left( z_i-1 \right) a_{{4}}.
\end{align}
In the sequel we will use the following notation, $x_{ij}=x_ix_j$, $c_{ij}=c_ic_j$, and
if $Q_i=Q(x_i,y_i,z_i)$ then $Q_{ij}=Q(x_{ij},y_{ij},z_{ij})$.
\subsection*{2-soliton}
Equation (\ref{DAKP}) admits the 2-soliton solution
\[
\tau_{k,l,m}=1+c_1x_1^ky_1^lz_1^m+c_2x_2^ky_2^lz_2^m+c_1c_2 R_{12}x_{12}^ky_{12}^lz_{12}^m
\]
where $Q_1=Q_2=0$,
\[
R_{ij}=
\frac{a_1S^{ij}_1+a_2S^{ij}_2+a_3S^{ij}_3+a_4S^{ij}_4}{Q_{ij}},
\]
with
\begin{align*}
S^{ij}_1=\bigg( &
\left( x_i-x_j \right)  \left( x_jy_i-y_jx_i \right)  \left( x_{ij}-z_{ij} \right)
+ \left( x_i-x_j \right)  \left( x_jz_i-z_jx_i \right) \left( x_{ij}-y_{ij} \right)\\
&+ \left( x_jz_i-z_jx_i \right)  \left( x_jy_i-y_jx_i \right)  \left( 1-x_{ij} \right)
\bigg)y_{ij}z_{ij},\\
S^{ij}_4=\bigg( &
\left(1 - x_{ij} \right) \left( y_i-y_j \right) \left( z_i-z_j \right)
+ \left( x_i-x_j \right) \left(1 - y_{ij} \right) \left( z_i-z_j \right) \\
&+ \left( x_i-x_j \right) \left( y_i-y_j \right) \left(1 - z_{ij} \right)
\bigg)x_{ij}y_{ij}z_{ij},
\end{align*}
and $S^{ij}_k$ for $k=2$, resp. $k=3$, are obtained from $S^{ij}_1$ by interchanging the symbols $x$ and $y$, respectively $x$ and $z$. This has been checked by direct computation, using a Groebner basis in Maple \cite{MAP}.

\subsection*{$N$-soliton}
Let $P(N)$ denote the powerset of the string $12\ldots N$, e.g. we write
\[
P(3)=\{\varepsilon ,1,2,3,12,23,13,123\},
\]
where $\varepsilon$ is the empty string, and let $P_2(S)$ be the subset of the powerset of a string $S$ containing all 2-letter substrings, e.g.
\[
P_2(123)=\{12,23,13\}.
\]

\begin{conjecture}
Equation (\ref{DAKP}) admits the following $N$-soliton solution:
\[
\tau_{k,l,m}=\sum_{w \in P(N)} \Big(\prod_{v\in P_2(w)}R_v\Big) c_w x_w^k y_w^l z_w^m,\qquad \text{with } Q_i=0,\  i\in\{1,2,\ldots,N\}.
\]
\end{conjecture}
Note that in the above formula $c_\varepsilon=x_\varepsilon=\cdots=1$ is understood. The formula can be computationally checked as follows: Taking particular values for $a_1,a_2,a_3$ and $a_4$, one can find rational points $p_i=(x_i,y_i,z_i)\in\mathbb{Q}^3$ such that $Q_i=0$. Using $N\in\mathbb{N}$ points, one substitutes the $N$-soliton solution, which contains $N$ arbitrary constants $c_1,\ldots,c_N$, into the equation for fixed points $(k,l,m)\in\mathbb{Z}^3$. For example, taking $(a_1,a_2,a_3,a_4)=(1,2,3,2)$ the following points
\begin{align*}
&p_1=(2, 4, 2/3),\
&&p_2=(6, 21, -14),\
&&p_3=(7, 14, -6),\\
&p_4=(8, 15, -40/9),\
&&p_5=(14, 80, -560),\
&&p_6=(18, 120, -15/2)
\end{align*}
satisfy $Q_i=0$. Taking $k=-2,l=1,m=3$ one needs to verify that
\begin{equation}\label{equ}
\begin{split}
&\tau_{{-3,2,4}}\tau_{{-1,1,3}}\tau_{{-1,1,4}}\tau_{{-1,2,3}}+2\,\tau_{{-2,1,3}}\tau_{{-2
,2,4}}\tau_{{-1,1,4}}\tau_{{-1,2,3}}-\tau_{{-2,1,4}}\tau_{{-2,2,3}}\tau_{{-2,2,4}}\tau_{
{0,1,3}}\\
&-2\,\tau_{{-2,1,4}}\tau_{{-2,2,3}}\tau_{{-1,1,3}}\tau_{{-1,2,4}}+3\,\tau_{{-2
,1,4}}\tau_{{-2,2,4}}\tau_{{-1,1,4}}\tau_{{-1,2,2}}-2\,\tau_{{-2,1,4}}\tau_{{-2,3,3}}
\tau_{{-1,1,3}}\tau_{{-1,1,4}}\\
&-3\,\tau_{{-2,1,5}}\tau_{{-2,2,3}}\tau_{{-1,1,3}}\tau_{{-1
,2,3}}+2\,\tau_{{-2,2,3}}\tau_{{-2,2,4}}\tau_{{-1,0,4}}\tau_{{-1,2,3}}
\end{split}
\end{equation}
vanishes. Using the above 6 points $p_i$ the value of the 6-soliton solution at
$(k,l,m)=(-3,2,4)$ is
{\small
\begin{align*}
\tau_{{-3,2,4}}&=1+{\frac {32\,c_{{1}}}{81}}+{\frac {235298\,c_{{2}}}{3
}}+{\frac {5184\,c_{{3}}}{7}}+{\frac {125000\,c_{{4}}}{729}}-{\frac {
9973408256.10^{8}\,c_{{1}}c_{{2}}c_{{3}}c_{{4}}}{2850829229061}} +{
\frac {17537436614656.10^{14}\,c_{{1}}c_{{4}}c_{{5}}}{
304882184692881}}\\
&-{\frac {286643773308928.10^{13}\,c_{{2}}c_{{4}}
c_{{5}}}{5379614362287}}
-{\frac {31023435087872.10^{14}\,c_{{3}}
c_{{4}}c_{{5}}}{1827893357279451}}-{\frac {8355684882055168.10^{7}\,c_
{{1}}c_{{2}}c_{{5}}}{3557331}}-{\frac {244.10^{6}\,c_{{1}}c_{{4}}}{2735937}}\\
&+{\frac {419082155327488.10^{10}\,c_{{
1}}c_{{3}}c_{{5}}}{79592065203}}-{\frac {25192657019901837312.10^{7}\,
c_{{2}}c_{{3}}c_{{5}}}{414577637413}}+{\frac {15625\,c_{{6}}}{2}}+
229376.10^{6}\,c_{{5}}+{\frac {430515.10^{5}\,c_{{3}}c_{{6}}}{313747}}\\
&+{
\frac {1220703125\,c_{{4}}c_{{6}}}{34992}}-{\frac {6272.10^{13}\,
c_{{5}}c_{{6}}}{339}}+{\frac {5.10^{5}\,c_{{1}}c_{{6}}}{1539}}-{\frac {
1838265625\,c_{{2}}c_{{6}}}{422}}+{\frac {
4563788408614224681500672.10^{17}\,c_{{1}}c_{{2}}c_{{3}}c_{{4
}}c_{{5}}}{3335745327609453768757318923}}
\end{align*}}
{\small
\begin{align*}
\phantom{\tau_{{-3,2,4}}}
&-{\frac {1075648.10^{6}\,c_{{2
}}c_{{3}}c_{{4}}}{253204479}}+{\frac {9415684768.10^{6}\,c_{{1}}c_{{2}}
c_{{4}}}{1595051271}}-{\frac {74176.10^{8}\,c_{{1}}c_{{3}}c_{{4}}}{
10556764911}}+{\frac {359696691200\,c_{{1}}c_{{2}}c_{{3}}}{13161}}-{\frac {323830284288.10^{6}\,
c_{{2}}c_{{5}}}{367}}\\
&+{
\frac {68956750243187158016.10^{25}\,c_{{1}}c_{{2}}c_
{{3}}c_{{4}}c_{{5}}c_{{6}}}{13520106051588549460696520569492377}}+{
\frac {203190312540790063104.10^{17}\,c_{{1}}c_{{2}}c_{{3}}c_
{{5}}c_{{6}}}{22721904043520272815643}}-{\frac {270945647.10^{21}\,c_{{2}}
c_{{4}}c_{{5}}c_{{6}}}{3405295891327671}}\\
&+{\frac {
2558749998443072.10^{22}\,c_{{1}}c_{{2}}c_{{4}}c_{{5}}c_
{{6}}}{87172894850121902722923}}+{\frac {
512857367787136.10^{25}\,c_{{1}}c_{{3}}c_{{4}}c_{{5}}
c_{{6}}}{17456412275319361592913150717}}-{\frac {5065859375.10^{8}
\,c_{{1}}c_{{3}}c_{{4}}c_{{6}}}{1419494280227793}}-{
\frac {8.10^{6}\,c_{{3}}c_{{4}}}{22869}}\\
&+{\frac {
1197498441289024.10^{22}\,c_{{2}}c_{{3}}c_{{4}}c_{{5}}c_
{{6}}}{3344920733568156174088032717}}+{\frac {486525134375.10^{10}\,
c_{{1}}c_{{2}}c_{{3}}c_{{4}}c_{{6}}}{3851564365825970013}}+{\frac {
433061888.10^{7}\,c_{{1}}c_{{5}}}{29079}}-{\frac {
5190429687500\,c_{{3}}c_{{4}}c_{{6}}}{3075034347}}\\
&-{\frac {3785197879296.10^{7}\,c_{{3}}c_{{5}}}{
1802479}}-{\frac {152276992.10^{13}\,c_{{4}}c_{{5}}}{226286703}}-
{\frac {23683072.10^{14}\,c_{{1}}c_{{5}}c_{{6}}}{187297839}}
+{\frac {1353499793462147416064.10^{14}\,c_{{1}
}c_{{2}}c_{{4}}c_{{5}}}{7248099679897056849}}\\
&-{\frac {24118045.10^{5}\,c_{{1}}c_{{2}}c_{{6}}}{324729}}
+{\frac {34110994995.10^{5}\,c_{{2}}c_{{3}}c_{{6}}}{5926981771}}-{\frac
{297851562500\,c_{{1}}c_{{4}}c_{{6}}}{155948409}}+{\frac {
287875.10^{18}\,c_{{4}}c_{{5}}c_{{6}}}{2036580327}}+{\frac {
308710976\,c_{{1}}c_{{2}}}{243}}\\
&+{\frac {
143614501953125\,c_{{2}}c_{{4}}c_{{6}}}{358705908}}-{\frac {
8504.10^{8}\,c_{{1}}c_{{3}}c_{{6}}}{38590881}}-{\frac {
4427367168.10^{13}\,c_{{2}}c_{{5}}c_{{6}}}{8750381}}-{\frac {972800\,c_{{1}}c_{{3}}}{861}}-{\frac {224599144.10^{8}\,c_{{1}}c_{{2}}c_{{3
}}c_{{6}}}{5926981771}}\\
&-{\frac {1916804736\,c_{{
2}}c_{{3}}}{4387}}-{\frac {117649.10^{6}\,c_{{2}}c_{{4}}}{425007}}+{\frac {
929743819565759987712.10^{10}\,c_{{1}}c_{{2}}c_{{3}}c_{{5}}}{
148833371831267}}+{
\frac {5239147585536.10^{14}\,c_{{3}}c_{{5}}c_{{6}}}{
1304163853181}}\\
&-{\frac {
16971308117950201856.10^{17}\,c_{{1}}c_{{3}}c_{{4}}c_{{5}}}{
302920719026139857929671}}+{\frac {8342573976349835264.10^{14}\,
c_{{2}}c_{{3}}c_{{4}}c_{{5}}}{825274865866379324583}}-{\frac {
183176122990592.10^{17}\,c_{{1}}c_{{3}}c_{{5}}c_{{6}}}{
172763889794740251}}\\
&-{\frac {66307976.10^{22}\,c_{{1}}c_
{{4}}c_{{5}}c_{{6}}}{52134853582482651}}+{\frac {4984888671875.10^{4}\,c_
{{2}}c_{{3}}c_{{4}}c_{{6}}}{342087606876807}}-{\frac {
52304285961241509888.10^{14}\,c_{{2}}c_{{3}}c_{{5}}c_{{6}}}{
63292211820390732077}}\\
&+{\frac {
8906247704.10^{22}\,c_{{3}}c_{{4}}c_{{5}}c_{{6}}}{
105336010499942922777}}-{\frac {574687791015625.10^{4}\,c_{{1}}c_{{2}}c_{
{4}}c_{{6}}}{6394560545439}}-{\frac {228475758493696.10^{14}\,c_
{{1}}c_{{2}}c_{{5}}c_{{6}}}{1611531417627}},
\end{align*}
}

\noindent
and we obtain similar expressions for the values of the 6-soliton solution
at the other 13 lattice points of the stencil, cf. Figure \ref{STC}. Substituting these
expressions into (\ref{equ}) gives 0. This has been checked also for values of $a_i$,
other point $p_j$ and other values for $k,l,m$.

We have also performed another computational verification, this time of the 3-soliton solution. Starting with expressions for $p_1,p_2,p_3$ of the form $p_i=b_ix+c_i$ where $b_i,c_i\in\mathbb{Q}$
are randomly chosen and $x$ is a parameter, we have solved the linear system $Q_{12}=Q_{13}=Q_{23}=0$
for $a_1,a_2,a_3,a_4$, and verified the solution for a range of values for $k,l,m$.

\begin{figure}[H]
\begin{center}
\includegraphics[width=6.6cm]{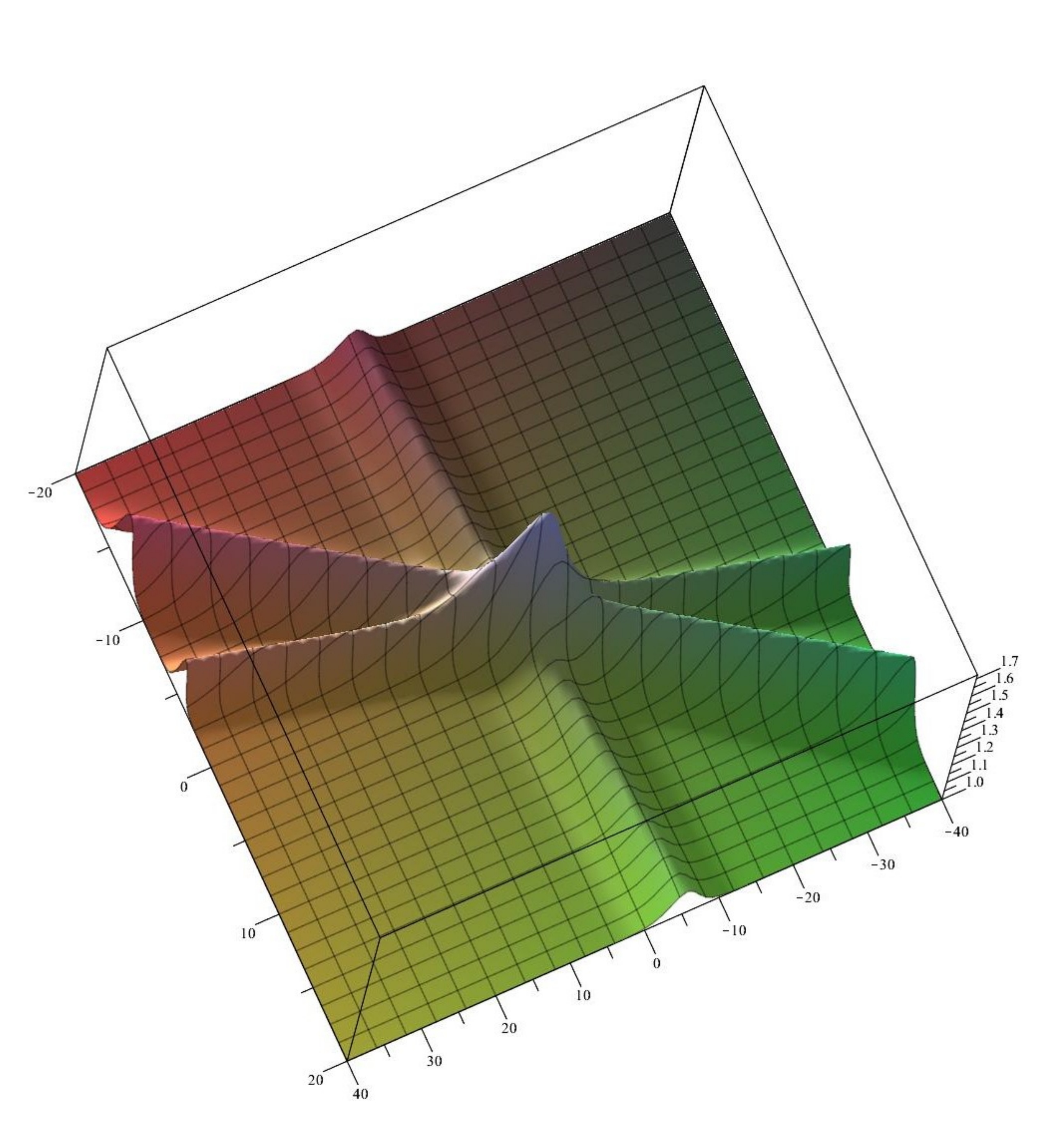}\hspace{1.5cm}
\includegraphics[width=6.6cm]{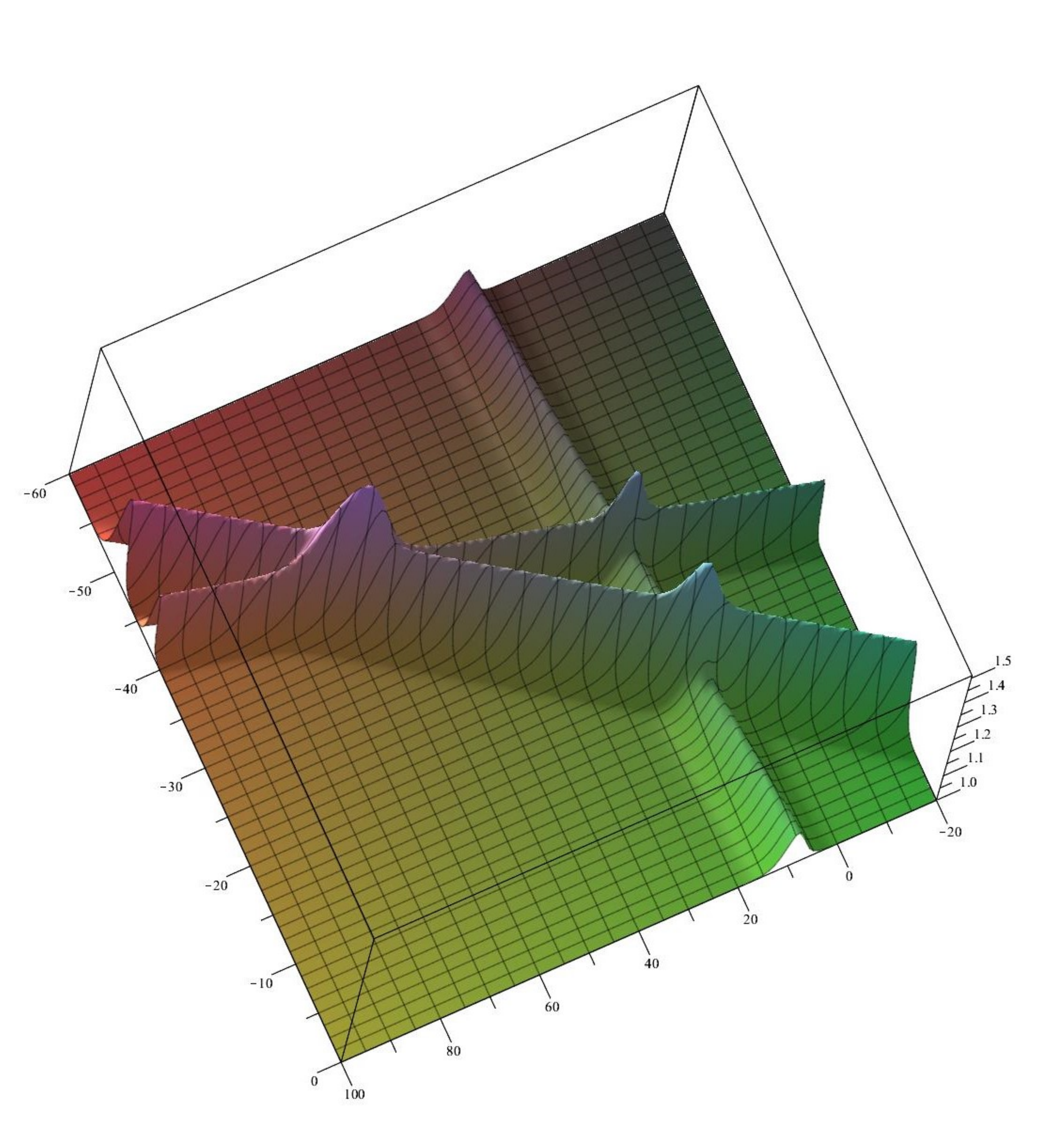}
\caption{\label{sol} Two cross sections, $m=0$ resp. $m=50$, of the function $u$ defined in (\ref{uzwv}) where $\tau$ is the three soliton solution of dual AKP with $(a_1,a_2,a_3,a_4)=(1,2,3,2)$ and $p_1=(\frac 15, \frac{12}{13}, \frac{18}{65})$, $p_2=(\frac{11}{31}, \frac{15}{32}, \frac{495}{496})$, $p_3=(2, 4, \frac 23)$, and $c_1=c_2=c_3=1$.}
\end{center}
\end{figure}

In Figure \ref{sol} we have plotted two cross sections of a three soliton solution.

\section{Laurent property} \label{sLp}
Consider an ordinary difference equation of order $d$,
\begin{equation}\label{IM1}
\tau_n=\frac{P(\tau_{n-d},\dots,\tau_{n-1})}{Q(\tau_{n-d},\dots,\tau_{n-1})},
\end{equation}
where  $P$ is a polynomial and $Q$ is a monomial. Let ${\cal R}$ be the ring of coefficients.
From a set of $d$ initial values $U=\{\tau_{k}\}_{0\leq k< d}$, one finds $\tau_{n}$ as  rational functions of the initial values, given by
\begin{align}\label{IM2}
\tau_{n}=\frac{p_{n}(\tau_{0},\dots,\tau_{d-1})}{q_{n}(\tau_{0},\dots,\tau_{d-1})},
\end{align} with greatest common divisor gcd$(p_{n},q_{n})=1$.  By definition, if $ q_{n}\in{\cal R}[U]$ is a monomial for all $n\geq0$, then (\ref{IM1}) has the Laurent property. The first examples of recurrences with the Laurent property were discovered by Michael Somos in the 1980s \cite{Gal}. Since then many more have been found \cite{ACH,EZ,FZ,HK,LP}, and the Laurent property is a central feature of cluster algebras \cite{fziv, fst}. In \cite[Definition 2.11]{Mas} the author defines the Laurent property for discrete bilinear equations. The idea is that a lattice equation has the Laurent property if all {\em good} initial value problems have the Laurent property. The author points out that not all well-posed, cf. \cite{vdK}, initial value problems are good. Certainly, the initial value problems obtained from (doubly periodic) reductions given below, see (\ref{LAKP}), are good.

In \cite{HHKQ} a more specific Laurent property was introduced, where the terms are Laurent polynomials in some of the variables but polynomial in others. The form of (\ref{IM1}) guarantees that all components $q_{n}$ are monomials for $0\leq n \leq d$. Suppose these monomials depend on a subset of the initial values $V\subset U$, specified by a set of superscripts $I\subset \{1,\ldots,d\}$. The following conditions guarantee that $q_{n}$ is a monomial $\in {\cal R}[V]$ for all $i$ and $n\geq0$, cf. \cite[Theorem 2]{HHKQ}.

\begin{theorem} \label{THM}
Suppose that $q_{d}$ is a monomial in ${\cal R}[V]$.
If $p_{d}$ is  coprime to $p_{d+k}$ for all $k =1,\dots,d$, and $q_{m}\in {\cal R}[V]$ is a monomial for $d+1\leq m \leq 2d$, then (\ref{IM1}) has the following Laurent property: all iterates are Laurent polynomials in the variables from $V$ and they are polynomial in the remaining variables from $W=U\setminus V$.
\end{theorem}

Introducing the variable $n=z_1k+z_2l+z_3m$, where we take $z_1,z_2,z_3$ to be non-negative integers such that gcd$(z_1,z_2,z_3)=1$, and performing a reduction $\tau_{k,l,m}\rightarrow \tau_n$, one obtains
the ordinary difference equation
\begin{equation} \label{LAKP}
\begin{split}
0&=a_{{1}} \left(
\tau_{{n+z_{{1}}}}\tau_{{n+z_{{1}}+z_{{2}}}}\tau_{{n+z_{{1}}+z_{{3}}}}\tau_{{n-z_{{1}}+z_{{2}}+z_{{3}}}}
-\tau_{{n+2\,z_{{1}}}}\tau_{{n+z_{{2}}}}\tau_{{n+z_{{3}}}}\tau_{{n+z_{{2}}+z_{{3}}}} \right)\\
&\ \ \ +a_{{2}} \left(
\tau_{{n+z_{{2}}}}\tau_{{n+z_{{1}}+z_{{2}}}}\tau_{{n+z_{{2}}+z_{{3}}}}\tau_{{n+z_{{1}}-z_{{2}}+z_{{3}}}}
-\tau_{{n+z_{{1}}}}\tau_{{n+2\,z_{{2}}}}\tau_{{n+z_{{3}}}}\tau_{{n+z_{{1}}+z_{{3}}}}\right)\\
&\ \ \ +a_{{3}} \left(\tau_{{n+z_{{3}}}}\tau_{{n+z_{{1}}+z_{{3}}}}\tau_{{n+z_{{2}}+z_{{3}}}}\tau_{{n+z_{{1}}+z_{{2}}-z_{{3}}}} -\tau_{{n+z_{{1}}}}\tau_{{n+z_{{2}}}}\tau_{{n+2\,z_{{3}}}}\tau_{{n+z_{{1}}+z_{{2}}}}\right)\\
&\ \ \ +a_{{4}} \left( \tau_{{n}}\tau_{{n+z_{{1}}+z_{{2}}}}\tau_{{n+z_{{1}}+z_{{3}}}}\tau_{{n+z_{{2}}+z_{{3}}}}
-\tau_{{n+z_{{1}}}}\tau_{{n+z_{{2}}}}\tau_{{n+z_{{3}}}}\tau_{{n+z_{{1}}+z_{{2}}+z_{{3}}}}
\right),
\end{split}
\end{equation}
which has order
\[
d=\max(2z_1,2z_2,2z_2,z_1+z_2+z_3)-\min(0,z_1+z_2-z_3,z_1+z_3-z_2,z_2+z_3-z_1).
\]

\begin{conjecture}
The iterates $\tau_n$ are Laurent polynomials in the initial values $\tau_{i}$, with $i=p,p+1,\ldots,d-p-1$ where
\[
p=\min(z_1,z_2,z_2)-\min(0,z_1+z_2-z_3,z_1+z_3-z_2,z_2+z_3-z_1),
\]
and polynomial in the others, $\tau_0,\tau_1,\ldots,\tau_{p-1},\tau_{d-p},\ldots,\tau_{d-2},\tau_{d-1}$.
\end{conjecture}

This conjecture has been proven, using Theorem \ref{THM} and \cite{MAP}, for $z_1=z_2=1$, $1\leq z_3\leq 20$, for $z_1=1,z_2=2,z_3=3$, and some but not all of the conditions of Theorem \ref{THM} have been verified for all co-prime $z_1<z_2<z_3\leq 10$.

\section{Degree growth} \label{DG}
Given an ordinary difference equation of the form
(\ref{IM1}) one can define an integer sequence $\{d^p_n\}_{n=0}^\infty$ where $d^p_n$ denotes the degree of the polynomial $p_n$ defined by (\ref{IM2}).
According to the {\em degree growth conjecture} \cite{FV,HV} we have
\begin{itemize}
\item growth is linear in $n$ $\implies$ equation is linearizable.
\item growth is polynomial in $n$ $\implies$ equation is integrable.
\item growth is exponential in $n$ $\implies$ equation is non-integrable.
\end{itemize}

\begin{conjecture}
For all positive integers $z_1,z_2,z_3$ such that gcd$(z_1,z_2,z_3)=1$ equation (\ref{LAKP}) has quadratic growth.
\end{conjecture}

We have verified the following. Choosing (randomly) rational values for the coefficients $a_i$, starting with rational initial values $\tau_0,\ldots,\tau_{d-2}$ and letting $\tau_{d-1}=a+bx$, where $a,b$ are rational values and $x$ a parameter, we have calculated up to a 150 iterates until the degree (in $x$) exceeded 250. Taking the second difference of the degree sequence yielded a periodic sequence in almost all cases with $1\leq z_1\leq 4$, $1\leq z_2 \leq z_3 \leq 7$. In two cases more iterations were required. Keeping the maximal degree fixed at 500, for $z=(1,1,7)$ we calculated 370 iterations and found that the period of the second difference is 259, for $z=(3,7,7)$ we calculated 354 iterations and found that the period of the second difference is 240. Curiously, the leading order terms are all of the form $(M^{z_1}_{z_2,z_3})^{-1} n^2$ with
\begin{align*}
M^1&=\left[ \begin {array}{ccccccc} 2&4&15&40&85&156&259
\\ \noalign{\medskip}4&7&12&25&60&94&172\\ \noalign{\medskip}15&12&16&
24&40&76&150\\ \noalign{\medskip}40&25&24&29&40&60&108
\\ \noalign{\medskip}85&60&40&40&46&60&82\\ \noalign{\medskip}156&94&
76&60&60&67&84\\ \noalign{\medskip}259&172&150&108&82&84&92
\end {array} \right],\qquad
M^2=\left[ \begin {array}{ccccccc} 4&7&12&25&60&94&172
\\ \noalign{\medskip}7&x&15&x&40&x&154\\ \noalign{\medskip}12&15&28&25
&60&55&132\\ \noalign{\medskip}25&x&25&x&40&x&76\\ \noalign{\medskip}
60&40&60&40&84&60&140\\ \noalign{\medskip}94&x&55&x&60&x&82
\\ \noalign{\medskip}172&154&132&76&140&82&172\end {array} \right],
\\
M^3&=\left[ \begin {array}{ccccccc} 15&12&16&24&40&76&150
\\ \noalign{\medskip}12&15&28&25&60&55&132\\ \noalign{\medskip}16&28&x
&40&40&x&77\\ \noalign{\medskip}24&25&40&69&60&55&168
\\ \noalign{\medskip}40&60&40&60&114&76&76\\ \noalign{\medskip}76&55&x
&55&76&x&108\\ \noalign{\medskip}150&132&77&168&76&108&240\end {array}
 \right],\qquad
M^4=\left[ \begin {array}{ccccccc} 40&25&24&29&40&60&108
\\ \noalign{\medskip}25&x&25&x&40&x&76\\ \noalign{\medskip}24&25&40&69
&60&55&168\\ \noalign{\medskip}29&x&69&x&85&x&77\\ \noalign{\medskip}
40&40&60&85&136&94&132\\ \noalign{\medskip}60&x&55&x&94&x&150
\\ \noalign{\medskip}108&76&168&77&132&150&296\end {array} \right],
\end{align*}
where $x$ indicates that gcd$(z_1,z_2,z_3)>1$.

\section{Reductions to 2D integrable lattice equations} \label{srt2dle}
We give some reductions to integrable 2D lattice equations known in the literature.
\begin{itemize}
\item
Setting\ \ $\dot{ }=\hat{ }$ , $u=e$, $v=\tilde{q}$, $z=w=0$ and $a_1+a_4=0$ the Q$^3$D-system reduces
to Rutishauser's QD-algorithm
\[
\tilde{e}+\tilde{q}=\hat{e}+\tilde{\hat{q}},\qquad \hat{e}\hat{q}=e\tilde{q}.
\]
\item
Taking $z=a_2=0$ and $a_1=a_3=-a_4=1$ and introducing variables $i=k-l$,
$j=3l+m$, $\tau_{k,l,m}=\Delta_i^j$, equation (\ref{DAKP}) reduces to the higher analogue of the discrete-time Toda (HADT) equation \cite[Equation (3.18)]{SNK},
\begin{align*}
&\Delta_{{i+1}}^{{j}} \left( \Delta_{{i-2}}^{{j+4}}\Delta_{{i+1}}^{{j+1}}\Delta_{{i
}}^{{j+3}}-\Delta_{{i}}^{{j+2}}\Delta_{{i-1}}^{{j+3}}\Delta_{{i}}^{{j+3}}+\Delta_{{i-1
}}^{{j+3}}\Delta_{{i}}^{{j+1}}\Delta_{{i}}^{{j+4}} \right)\\
&=\Delta_{{i-1}}^{{j+4}}
 \left( -\Delta_{{i+2}}^{{j}}\Delta_{{i}}^{{j+1}}\Delta_{{i-1}}^{{j+3}}+\Delta_{{i
}}^{{j+2}}\Delta_{{i}}^{{j+1}}\Delta_{{i+1}}^{{j+1}}-\Delta_{{i}}^{{j}}\Delta_{{i+1}}^{{j+
1}}\Delta_{{i}}^{{j+3}} \right),
\end{align*}
and the Q$^3$D-system reduces to the QQD-system \cite[Equation (1.4)]{SNK},
\begin{eqnarray*}
u_{{i,3+j}}+v_{{i+1,j+1}}+w_{{i+1,j}}&=&u_{{i+2,j}}+v_{{i+1,j}}+w_{{i+1,j+1}}\\
u_{{i,3+j}}v_{{i,j+1}}&=&u_{{i+1,j}}v_{{i+1,j}}\\
u_{{i,3+j}}w_{{i,j+1}}&=&u_{{i+1,j+1}}w_{{i+1,j+1}}.
\end{eqnarray*}
\item By introducing some special bi-orthogonal polynomials, in \cite{CCHT} the so-called discrete hungry quotient-difference (dhQD) algorithm and a system related to the QD-type discrete hungry Lotka-Volterra (QD-type dhLV) system have been derived, as well as hungry forms of the HADT-equation (hHADT) and the QQD scheme (hQQD). These systems are all reductions of the Q$^3$D system,
or of the dual to the AKP equation, (\ref{DAKP}).

Setting $z=w=0$, $\tilde{u}=q$, $v=e$ and introducing $i=k$, $j=pl+m$ we get QD-type dhLV \cite[Equations (6,7)]{CCHT},
\begin{eqnarray*}
e_{{i,j}}+q_{{i,j}}&=&e_{{i,j+1}}+q_{{i-1,j+p}}\\
e_{{i,j+1}}q_{{i,j+p}}&=&e_{{i+1,j}}q_{{i,j}}.
\end{eqnarray*}

Setting $z=w=0$, $\tilde{u}=q$, $v=e$ and introducing $i=k$, $j=l+pm$ we get dhQD \cite[Equations (9,10)]{CCHT},
\begin{eqnarray*}
e_{{i,j}}+q_{{i,j}}&=&e_{{i,j+p}}+q_{{i-1,j+1}}\\
e_{{i,j+p}}q_{{i,j+1}}&=&e_{{i+1,j}}q_{{i,j}}.
\end{eqnarray*}

With $z=0$ the reduction $i=k-l$, $j=(p+2)l+pm$ yield hQQD \cite[Equation (23)]{CCHT},
\begin{eqnarray*}
u_{{i,j+p+2}}+v_{{i+1,j+p}}+w_{{i+1,j}}&=&u_{{i+2,j}}+v_{{i+1,j}}+w_{{i+1,j+p}}\\
u_{{i,j+p+2}}v_{{i,j+p}}&=&u_{{i+1,j}}v_{{i+1,j}}\\
u_{{i,j+2}}w_{{i,j}}&=&u_{{i+1,j}}w_{{i+1,j}}.
\end{eqnarray*}

Performing the same reduction on equation (\ref{DAKP}), with $a_2=0$, gives the hHADT equation \cite[Equation (18)]{CCHT},
\begin{align*}
&\left( \Delta_{{i-2}}^{{j+2p+2}}\Delta_{{i+1}}^{{j+p}}\Delta_{{i}}^{{j+p+2}}
-\Delta_{{i}}^{{j+2p}}\Delta_{{i-1}}^{{j+p+2}}\Delta_{{i}}^{{j+p+2}}
+\Delta_{{i}}^{{j+2p+2}}\Delta_{{i}}^{{j+p}}\Delta_{{i-1}}^{{j+p+2}} \right) \Delta_{{i+1}}^{{j}}\\
&= \left(\Delta_{{i+2}}^{{j}}\Delta_{{i}}^{{j+p}}\Delta_{{i-1}}^{{j+p+2}}
-\Delta_{{i}}^{{j+2}}\Delta_{{i}}^{{j+p}}\Delta_{{i+1}}^{{j+p}}
+\Delta_{{i}}^{{j}}\Delta_{{i+1}}^{{j+p}}\Delta_{{i}}^{{j+p+2}} \right) \Delta_{{i-1}}^{{j+2p+2}}.
\end{align*}
\end{itemize}

\section{Conclusion}
In this paper we have generalized the concept of duality introduced in \cite{QCR}
for ordinary difference equations (O$\Delta$Es) to the realm of lattice equations
(P$\Delta$Es). The dAKP equation \eqref{DAKP} and the AKP equation \eqref{AKP} are dual to each other. Generally speaking, dual equations to integrable equations do not need to be integrable themselves; the only thing that is guaranteed is the existence of integrals (for O$\Delta$Es), or conservation laws (for P$\Delta$Es). However, our equation \eqref{AKP} unifies a number of known (hierarchies of) integrable 2D lattice equations, which arise as reductions. Together with the support we have provided for our conjectures, that equation \eqref{DAKP} admits an $N$-soliton solution, and its reductions have the Laurent property and zero algebraic entropy, we believe it is a new {\em integrable} 3D lattice equation.

\subsection*{Acknowledgements}
This research was supported by the Australian Research Council [DP140100383], by the NSF of China [No. 11371241, 11631007], and by two La Trobe University China Strategy Implementation Grants.

\end{document}